\documentclass[12pt]{iopart}

\usepackage{graphicx}

\newcommand{\et}{{{\it et al.}}}
\newcommand{\eg}{{{\it e.g.}}}
\begin{document}

\title[Parametric instabilities of Alfv\'en waves: 2-D PIC simulation]{Parametric instabilities of large-amplitude parallel propagating Alfv\'en waves: 2-D PIC simulation}

\author{Y Nariyuki$^1$$^2$, S Matsukiyo$^2$, and T Hada$^2$}

\address{$^1$Department of Earth System Science and Technology, Kyushu University, Japan\\
$^2$Department of Electrical Engineering, Kochi National College of Technology, Japan}
\ead{nariyuki@ee.kochi-ct.ac.jp}
\begin{abstract}

We discuss the parametric instabilities of large-amplitude parallel propagating Alfv\'en waves using the 2-D PIC simulation code. First, we confirmed the results in the past study [Sakai \et, 2005] that the electrons are heated due to the modified two stream instability and that the ions are heated by the parallel propagating ion acoustic waves. However, although the past study argued that such parallel propagating longitudinal waves are excited by transverse modulation of parent Alfv\'en wave, we consider these waves are more likely to be generated by the usual, parallel decay instability. Further, we performed other simulation runs with different polarization of the parent Alfv\'en waves or the different ion thermal velocity. Numerical results suggest that the electron heating by the modified two stream instability due to the large amplitude Alfv\'en waves is unimportant with most parameter sets.
\end{abstract}

\maketitle

\section{Introduction: file preparation and submission}
Large amplitude, low-frequency Alfv\'en waves constitute one of the most essential elements of magnetic fluctuations in the solar corona and solar wind\cite{tu95,bruno05}. Due to small collisionless dissipation rates, the waves can propagate long distance and efficiently convey such macroscopic quantities as momentum, energy, and helicity. Since loading of such quantities is completed when the waves damp away, it is important to examine how the waves can dissipate. Among various possible dissipation processes of the Alfv\'en waves, parametric instabilities have been believed to be important\cite{suzuki06,nariyuki07e}. Further, the parametric instabilities of these Alfv\'en waves can cause the developed turbulence, which are believed to produce \textit{in situ} the localized structures (shocks and discontinuities) in the solar wind\cite{tsurutani05,vasquez07}.

However, most past studies using kinetic simulation code mainly discussed the parametric instabilities among parallel propagating waves in one-dimensional system, although the real solar wind is three-dimensional. Actually, arguments including obliquely propagating waves have been developed by several authors using the fluid systems\cite{mjolhus90,vinaz91a,vinaz91b,ghosh93,ghosh94a,ghosh94b,laveder02}. These studies clarified that the growth rates involving the oblique waves are usually much smaller than those among parallel propagating waves.

On the other hand, Sakai \et\cite{sakai05} recently performed two-dimensional particle-in-cell (2-D PIC) simulations. They reported that the left-hand (LH-) polarized Alfv\'en waves are dissipated through the transverse modulational instability when ion and electron beta is very low ($=0.01$). This is in contrast to the past studies using the fluid systems which indicated that parametric instabilities including obliquely propagating daughter waves (transverse modulation or filamentation instability) can be dominant only in the high $\beta_{f}$ (the squared normalized sound velocity) plasma. One of the purposes of the present study is to clarify the reason of this difference on the type of parametric instabilities between the fluid and the PIC simulations.

Sakai \et\cite{sakai05} also reported the electron heating through the modified two stream instability (MTSI)\cite{matsukiyo06} which is caused by the difference in ion and electron bulk velocities (cross-field currents). In a presence of finite amplitude Alfv\'en waves, the cross-field currents can initially be given by the Walen relation. However, Sakai \et\cite{sakai05} only discussed LH- mode in their paper. Another purpose of this study is to confirm the importance of the MTSI driven by Alfv\'en waves with both right-hand (RH-) and left-hand (LH-) polarizations, and also by a mixture of Alfv\'en waves (in which the Walen condition is not satisfied). 

The paper is organized as follows. In section 2, we present our simulation model. In section 3, we present our simulation results. A summary and conclusions of this paper are given in section 4.

\section{Simulation setup}
We have carried out the 2-D PIC simulation\cite{matsukiyo06} with the system size $L_{x}=1024 - 2048$ and $L_{y}=512 - 1024$ with 100 super-particles (both ions and electrons) per cell. We use periodic boundary conditions in both $x$ and $y$ directions, with the uniform ambient magnetic field ($b_{x0}=1$) in the $x$-direction, and the homogeneous plasma (the zeroth order longitudinal bulk velocity $u_{xj0}=0$ ($j=i,e$) and the number density $n_{i0}=n_{e}=const$). The ion to electron mass ratio is $m_{i}/m_{e}=16$, and the ratio of the electron plasma to the cyclotron frequency is $\omega_{e}/|\Omega_{e}|=1$ ($\omega_{j}$ and $\Omega_{j}$ are plasma and the cyclotron angular frequencies of electrons ($j=e$) and ions ($j=i$), which include the ). Alfv\'en velocity $C_{A}=0.25c$, thus, when the beta ratio is $0.01$ for both ions and electrons, electron thermal velocity is $v_{th,e} (=\sqrt{2T_{e}/m_{e}}) =0.1c$, and ion thermal velocity is $v_{th,i} (=\sqrt{2T_{i}/m_{i}}) =0.025c$. The Debye length $v_{th,e}/\omega_{e}=1$ (grid size) and the electron skin depth $c/\omega_{e}=10$.
As initial conditions, finite amplitude, monochromatic Alfv\'en waves are given
\begin{eqnarray}
b_p = b_0 \exp(i(\omega_0 t -k_0 x)) , \label{eq32000}\\
u_{jp} = u_{j0} \exp(i(\omega_0 t -k_0 x)) , \label{eq32001}
\end{eqnarray}
where $b=b_{y}+ib_{z}$ (complex transverse magnetic field) and $u_{j}=u_{jy}+iu_{jz}$ (complex transverse bulk velocity of each species), $b_0=0.5$, $\omega_{0}$ and $k_{0}$ satisfy the dispersion relation of the parallel propagating electromagnetic waves in the two fluid (electron-ion) system\cite{baumjohann96},
\begin{equation}
\frac{k^{2}c^{2}}{\omega^{2}} = 1 - \frac{\omega_{e}^{2}}{\omega (\omega + \Omega_{e})} - \frac{\omega_{i}^{2}}{\omega (\omega + \Omega_{i})},  \label{eqa02}
\end{equation}
and $u_{j0}$ is given by the Walen relation\cite{hollweg93}
\begin{eqnarray}
u_{j0} = -{\omega_{0} \over k_{0}} {\Omega_{j} \over (\omega_{0}+\Omega_{j})}b_{0}. \label{eq32002}
\end{eqnarray}
We adopt the notation that the positive (negative) $\omega_{0}$ corresponds to the RH- (LH-) polarized waves. In our PIC simulation the charge neutrality cannot be assumed and displacement current can not be neglected due to the values assigned to $C_{A}$ and $\omega_{e}/|\Omega_{e}|$. While these parameters used in the present study are different from those in the past studies using the hybrid simulation code\cite{nariyuki07e,terasawa86,araneda07,nariyuki07d}, our results can basically be explained within the framework of the Hall-MHD theory as we will see later.


\begin{table}
\caption{\label{tab:table2}Parameters used in simulation runs. 
}
\begin{tabular}{cccccc}
Run & Polarization & $\omega_{0}/\Omega_i$ & $k_{0}C_A/\Omega_i$ & $v_{th,i}/c$ & $u_{j0}$\\
\hline
1 & L & $-0.503$ & $-0.736$ & $0.025$ & (\ref{eq32002})\\
2 & R & $0.56$ & $0.491$ & $0.025$ & (\ref{eq32002})\\
3 & (mixed) & (mixed) & $-0.736$ & $0.025$ & $u_{j0}=0$\\
4 & L & $-0.503$ & $-0.736$ & $0.1$   & (\ref{eq32002})\\
\end{tabular}
%

\end{table}

We performed four simulation runs with different parameters as tabulated in Table~\ref{tab:table2}. The system size is $L_{x}=1024$ and $L_{y}=512$ in Run 1, 3 and 4, and $L_{x}=2048$ and $L_{y}=512$ in Run2. We have confirmed that the results remain essentially the same when we extended the transverse system size to $L_{y}=1024$. Run 1 is almost the same as the run in Sakai \et\cite{sakai05}(LH- parent wave satisfying the Walen relation). In Run 2 the RH- parent waves are given, and in Run 3, magnetic and the velocity perturbations are given independently (and thus the Walen relation is not satisfied) in order to model the fluctuations at the chromosphere. In Run 4, we introduce higher ion temperature in order to examine the roles of kinetic modulational instability\cite{nariyuki07e,araneda07,nariyuki07d}. Initial Alfv\'en waves are all forward propagating in Runs 1, 2, and 4, while forward and backward propagating waves are mixed in Run 3.

To discuss the propagating direction of Alfv\'en waves, the two-dimensional ($x-y$) spacial Fourier transforming analysis were performed in this study. Since we focus on the low-frequency waves in this paper, the vertical and horizontal axes ($k_{y}$ and $k_{x}$) of the figures indicating the Fourier analysis (Fig.~\ref{3.2.3}, ~\ref{3.2.5}, ~\ref{3.2.7}) are normalized by the ion skin depth ($c/\omega_{i}=40$).

\section{Numerical results}

%
%
\begin{figure}
\noindent\includegraphics[width=28pc]{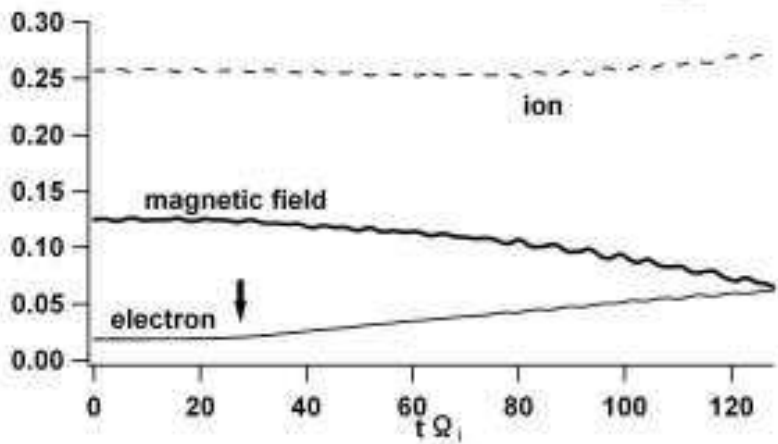}
\caption{\label{3.2.1} Time histories of magnetic field energy (thick line), ion (dashed line) and electron (thin line) total energies in Run 1.}
\end{figure}

\begin{figure}
\noindent\includegraphics[width=30pc]{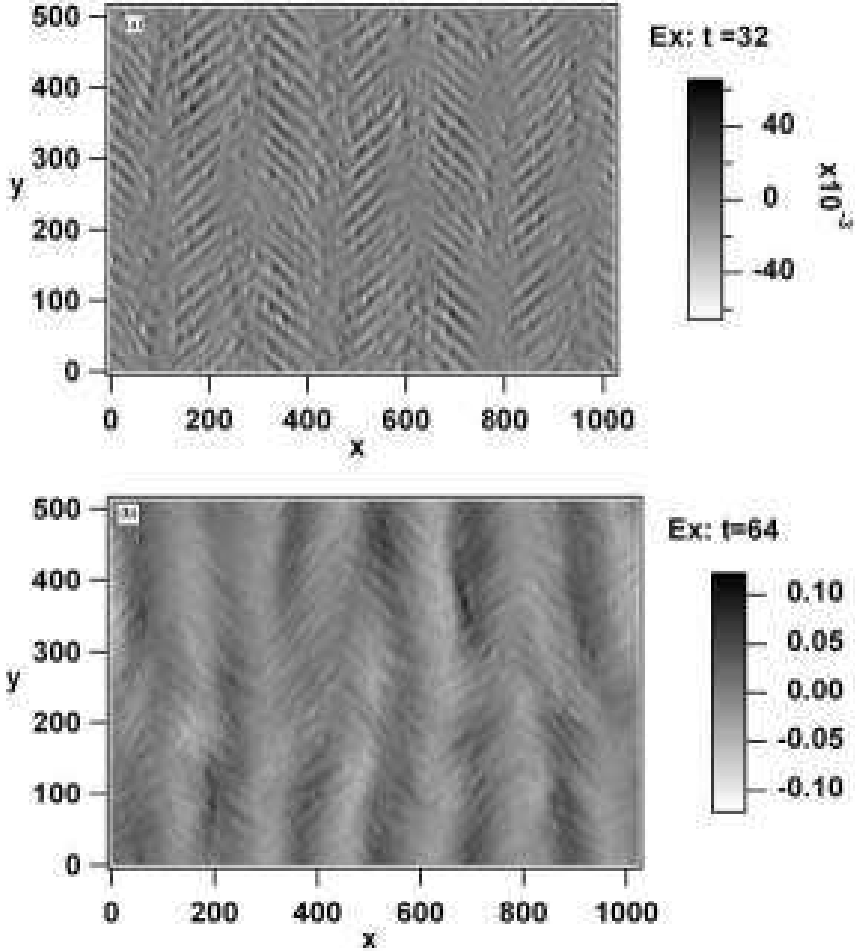}
\caption{\label{3.2.2} Snapshots of longitudinal electric field $e_{x}$ at (a)$t=32\Omega_{i}^{-1}$ and (b)$t=64\Omega_{i}^{-1}$ in Run 1.}
\end{figure}

First, we show the results of Run 1, in which physical parameters correspond to those in Sakai \et\cite{sakai05}. 
Fig.~\ref{3.2.1} shows the time history of magnetic field energy, ion and electron total energies, respectively. As discussed in Sakai \et\cite{sakai05}, the electron energy is enhanced by the excitation of MTSI (the arrow in Fig.~\ref{3.2.1}). Fig.~\ref{3.2.2}(a) clearly shows the obliquely propagating waves with parallel electric field, corresponding to Fig.3(b) in Sakai \et\cite{sakai05}. We have also confirmed that the propagation angle of the waves is about $60^{\circ}$, which agrees with the simulation results in Sakai \et \cite{sakai05} and the linear analysis in Wu \et \cite{wu83} (We note that while Wu \et \cite{wu83} discuss the case $\omega_{e}>>|\Omega_{e}|$, the growth rate of the MTSI is not sensitive to $\omega_{e}/|\Omega_{e}|$\cite{matsukiyo03}).The initial perpendicular velocity ($V_{0}$) is obtained from eq.(\ref{eq32002})
\begin{eqnarray}
V_{0}=(u_{i0}-u_{e0})=-v_{\phi 0}b_{0}\Omega_{i}{(1+r)\omega_{0} \over (\omega_{0}+\Omega_{j})(\omega_{0}-r\Omega_{j})}, \label{eq32004}
\end{eqnarray}
where $v_{\phi0}=\omega_{0}/k_{0}$, $r=m_{i}/m_{e}$. $V_{0}$ in Run 1 is $V_{0}=0.355 C_{A}$. We remark that the kinetic effects of electrons are important for the MTSI observed here, since $v_{th,e}>V_{0}$\cite{wu83}.

%
%

\begin{figure}
\noindent\includegraphics[width=28pc]{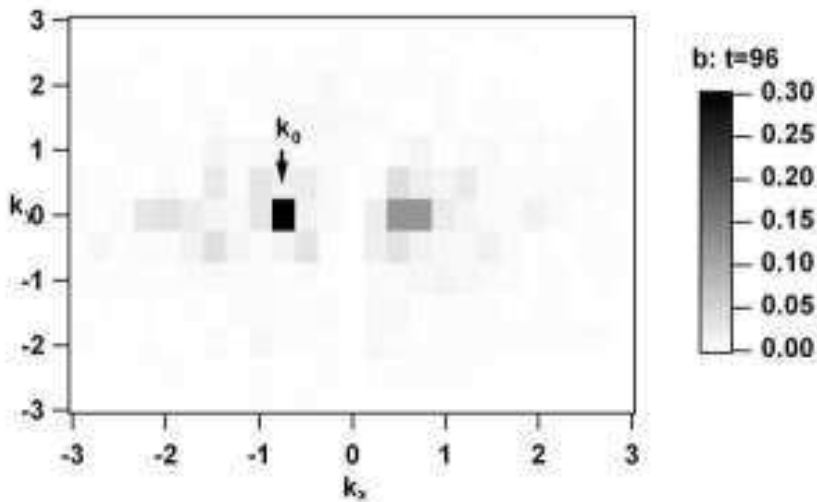}
\caption{\label{3.2.3} The power spectrum of the complex magnetic field $b$ ($=b_{y}+ib_{z}$) in $k_{x}-k_{y}$ wave number space at $t=96\Omega_{i}^{-1}$  in Run 1. The vertical and horizontal axes ($k_{y}$ and $k_{x}$) are normalized by the ion skin depth.}
\end{figure}

After these obliquely propagating waves excited by the MTSI are damped by electron Landau damping, low-frequency parallel propagating waves become dominant (Fig.~\ref{3.2.2}(b)). Dissipation of these parallel propagating ion acoustic waves heat the ions as discussed in Sakai \et\cite{sakai05}. Although Sakai \et\cite{sakai05} argued that such parallel propagating longitudinal waves are excited by transverse modulation of parent Alfv\'en wave, we consider these waves are more likely to be generated by the usual, parallel decay instability. Fig.~\ref{3.2.3} shows the power spectrum of the complex magnetic field $b$ ($=b_{y}+ib_{z}$) in the $k_{x}-k_{y}$ wave number space at $t=96\Omega_{i}^{-1}$. The anti-parallel propagating daughter waves ($\omega_{0}/k_{0}<0$) are dominantly excited around $(k_{x},k_{y})=(0.73,0)$. Presence of such anti-parallel propagating waves provides a strong evidence of the occurrence of the decay instability, while the daughter Alfv\'en waves excited by transverse modulation have finite propagation angles. In addition, these daughter waves are consistent with the linear analysis of Hall-MHD system for parallel propagating waves\cite{laveder02,terasawa86}. Thus, we conclude that under this set of parameters, the one-dimensional decay instability dominantly dissipates the parent Alfv\'en wave, as the Hall-MHD theory predicts\cite{laveder02}. While numerical results of the present study and Sakai \et\cite{sakai05} are quite similar, their interpretation on the parametric instability of parent Alfv\'en wave seems misleading. 

\begin{figure}
\noindent\includegraphics[width=28pc]{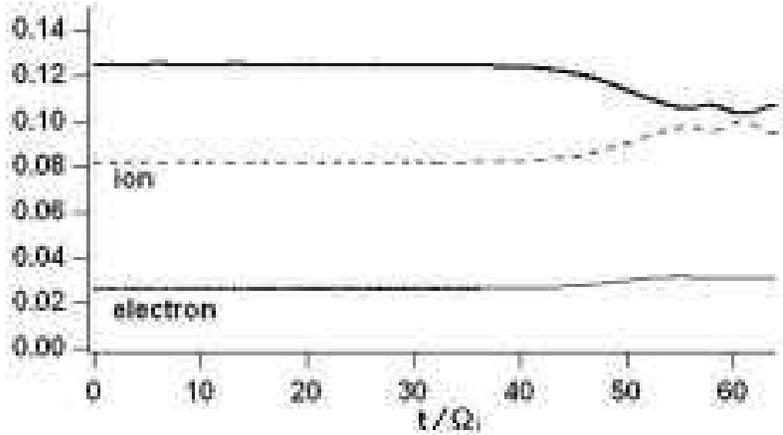}
\caption{\label{3.2.4} Time histories of magnetic field energy (thick line), ion (dashed line) and electron (thin line) total energies in Run 2.}
\end{figure}

\begin{figure}
\noindent\includegraphics[width=28pc]{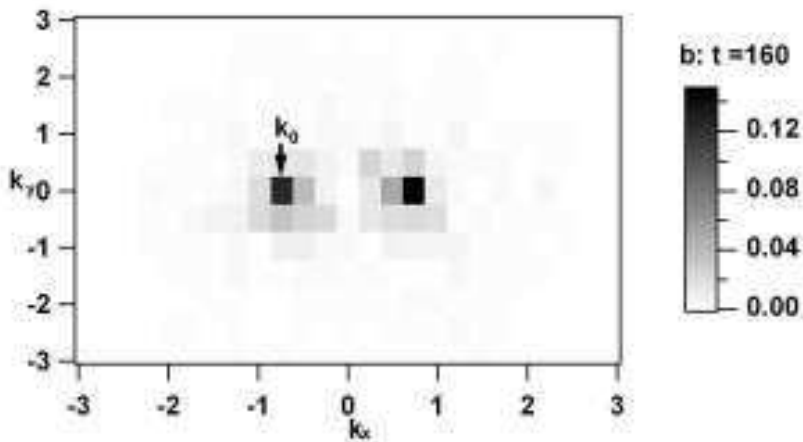}
\caption{\label{3.2.5} The power spectrum of the complex magnetic field $b$ ($=b_{y}+ib_{z}$) in $k_{x}-k_{y}$ wave number space at $t=160\Omega_{i}^{-1}$  in Run 3. The vertical and horizontal axes ($k_{y}$ and $k_{x}$) are normalized by the ion skin depth.}
\end{figure}

Next, we discuss the RH- mode case (Run 2). Fig.~\ref{3.2.4} shows the time history of energies in the same format as Fig.~\ref{3.2.1}. In contrast to Fig.~\ref{3.2.1} (Run 1), electron heating due to the excitation of MTSI is not evident. From eq. (\ref{eq32004}), $V_{0}$ in Run 2 is $0.223C_{A}$, which is larger than $v_{th,i}$. We here find that even when $V_{0}>v_{th,i}$, which is a rough criterion of the MTSI\cite{mcbride72}, the cross field current is not large enough to excite the MTSI under the particular set of parameters used. On the other hand, the decay instability takes place in a way similar to Run 1 (not shown).

%
%

%
%

In Run 3, we discuss the case in which the Walen relation is not satisfied for the initial Alfv\'en wave. These initial conditions are intended to simulate the plasma and the magnetic field fluctuations in the solar wind and solar surface, where both RH- and LH- polarized Alfv\'en waves are considered to be mixed\cite{bruno05}. In such a case, amplitude of transverse magnetic field and transverse bulk velocities vary in time\cite{nariyuki07d}. Since such linear oscillations modify the magnitude of the cross field current, they let the system stay in the stable and unstable regimes in turn, so that the MTSI growth is effectively reduced.

%
%

\begin{figure}
\noindent\includegraphics[width=28pc]{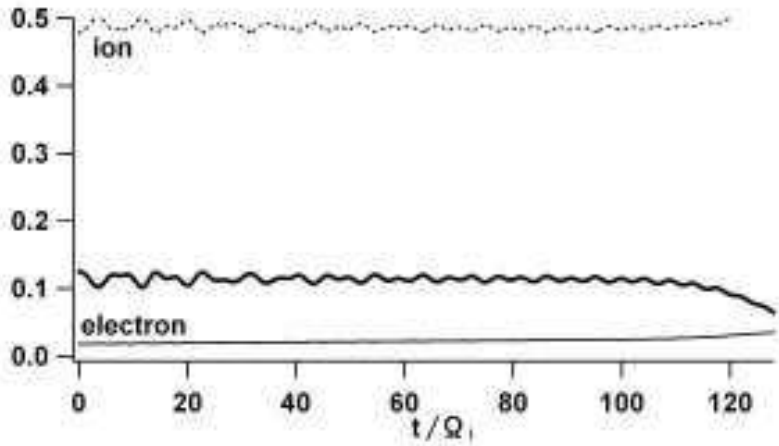}
\caption{\label{3.2.6} Time histories of magnetic field energy (thick line), ion (dashed line) and electron (thin line) total energies in Run 4.}
\end{figure}

\begin{figure}
\noindent\includegraphics[width=28pc]{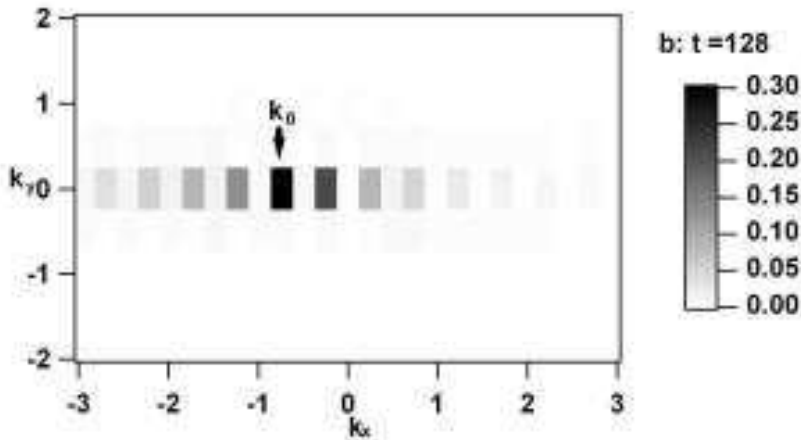}
\caption{\label{3.2.7} The power spectrum of the complex magnetic field $b$ ($=b_{y}+ib_{z}$) in $k_{x}-k_{y}$ wave number space at $t=128\Omega_{i}^{-1}$  in Run 4. The vertical and horizontal axes ($k_{y}$ and $k_{x}$) are normalized by the ion skin depth.}
\end{figure}

Finally, we discuss the results of Run 4, in which parameters are the same as those in Run 1 except for $v_{th,i}=0.1c$. In contrast to Run 1, MTSI is suppressed by the ion kinetic effects (Fig.~\ref{3.2.6}). Further, as Fig.~\ref{3.2.7} shows, the decay instability is not the dominant instability in this case, because of the occurrence of \lq\lq kinetic'' modulational instability\cite{nariyuki07d}. In Fig.~\ref{3.2.7}, both the dominant daughter waves ($k_{x} \sim -0.25, -1.25$) and the higher harmonic waves are observed.

%
%

%
%

\section{Summary}
In this paper, we discussed the parametric instabilities of Alfv\'en waves using a 2-D PIC simulation code. First, we re-ran the simulation of Sakai \et\cite{sakai05} using essentially the same set of parameters. While our numerical results are basically the same, we conclude that the generation of longitudinal waves are due to the usual, parallel decay instability, rather than the transverse modulational instability proposed by Sakai \et\cite{sakai05}, since the one-dimensional decay instability is dominant in a low beta plasma as the Hall-MHD theory predicted\cite{ghosh93,laveder02}. Further, we performed several other simulation runs using different parent Alfv\'en wave polarizations and different ion thermal velocities. The condition to excite the MTSI is easily violated by such modifications of the parameters. Thus, the electron heating by the MTSI of the Alfv\'en waves seems unimportant in the upper chromoshpere, since the Alfv\'en waves excited by the convection in the solar surface are likely to be the superposition of Alfv\'en waves with different polarizations and propagating directions, but not the monochromatic, LH- polarized wave. 
Further, a similar process on obliquely propagating waves have also been suggested by Markovskii and Hollweg \cite{markovskii02} according to the proton heating. The more detailed analysis on this issue is desired.

While the dominant parametric instabilities are quasi-one dimensional in a low beta case, the transverse modulational instability can be more important in different circumstances. For instance, the Alfv\'en wave excited by the ion beam instabilities can be dominantly dissipated through the transverse modulational instability, which is nonlinearly driven by the obliquely propagating waves excited by the same beam instabilities\cite{wang06b}. Further, even if the parametric instabilities are one dimensional, two or three dimensional density structures are excited as a consequence of its nonlinear evolution\cite{ghosh94a,delzanna01}. When the inhomoginity of the background plasmas exist, heating of electrons by phase mixed Alfv\'en waves take place\cite{tsiklauri05}. We remark that even in the uniform background plasmas, the electron heating regardless of the MTSI is actually observed in PIC simulation (\eg, Sakai \et\cite{sakai05} and  Fig.~\ref{3.2.4}). The electron heating is possibly caused by numerical noise originated from the super-particles in the PIC simulation. To avoid this and confirm the recent results, the electromagnetic Vlasov simulations are in plan.

\section*{Acknowledgments}
This paper has been supported by JSPS Research Fellowships for Young Scientists in Japan.

\end{document}